\documentclass[pre,amssymb,aps,floatfix]{revtex4-2}
\usepackage{graphicx,epstopdf}
\usepackage{dcolumn}
\usepackage[sort&compress]{natbib}
\usepackage{bm}
\usepackage[dvips]{epsfig}

\usepackage{amsmath}
\usepackage{xcolor} 
\usepackage[normalem]{ulem}

\newcommand{\comment}[1]{}

\newcommand{\BEA}{\begin{eqnarray}}
\newcommand{\EEA}{\end{eqnarray}}
\renewcommand{\r}{\bm{r}}
\newcommand{\m}{\nu_0}
\newcommand{\Nhb}{N_{\rm Hb}}
\renewcommand{\O}{${\rm O_2}$\,}

\newcommand{\mic}{\mu{\rm m}}

\begin{document}

\title{
Canonic hemoglobin-oxygen equilibrium: Reassessing the role of cooperativity  }

\author{A.E. Allahverdyan$^{1}$} 
\thanks{Corresponding author; e-mail: a.allahverdyan@aanl.am}
\author{S.G. Gevorkian$^{1,4}$, A. Harutyunyan$^{2}$, Y.Sh. Mamasakhlisov$^{2,3}$, V. Stepanyan$^{2}$}
\affiliation{$^{1}$Alikhanyan National Laboratory (Yerevan Physics
Institute), Yerevan, Armenia\\
$^{2}$Yerevan State University, Yerevan, Armenia\\
$^{3}$Institute of Applied Problems of Physics, Yerevan, Armenia\\
$^{4}$Institute of Physics, Academia Sinica, Nankang, Taipei, Taiwan\\
}

\begin{abstract} 

Hemoglobin-oxygen equilibrium is normally studied within the grand canonical ensemble, which assumes that each hemoglobin is immersed in a reservoir of unbound oxygen molecules of the plasma. We show that this assumption is incorrect inside the RBC (red blood cell or erythrocyte), where the hemoglobin concentration is larger than the concentration of unbound oxygen in plasma. We suggest a better model, where a single hemoglobin and a few oxygen molecules around it reach a canonical equilibrium at a fixed volume (determined from the RBC structure) and a fixed temperature. The basic models of hemoglobin-oxygen equilibrium -- Pauling's model and Monod-Wyman-Changeux model -- can be reformulated for this canonical situation. They predict cooperative interaction energies, which are significantly lower than predictions of the same models in the grand canonical ensemble. Larger cooperative energies, in particular those predicted by the grand canonical ensemble, lead to instabilities (cascade processes) in oxygen release within the canonical approach. Oxygen-binding fluctuations within this approach are sizably smaller than those in the grand canonical situation. These results suggest that the hopping diffusion of oxygen from one hemoglobin to another may play a role in oxygen diffusion within the RBC.

\end{abstract}

\maketitle

\section{Introduction}
\label{intro}

The hemoglobin molecule (Hb) plays several important physiological roles, including oxygen 
(\O) transport \cite{dill,popel}. Hb is also a key example of cooperative interactions in biological physics \cite{dill,popel}. Two major scenarios of cooperativity, induced fit and (shifted) conformational equilibrium, were developed in the context of \O binding to Hb \cite{pauling_pnas,eaton_models,mwc,changeux_review_50,koshland,ising}. 

\begin{figure}[t]
    \centering
    \includegraphics[width=0.55\linewidth]{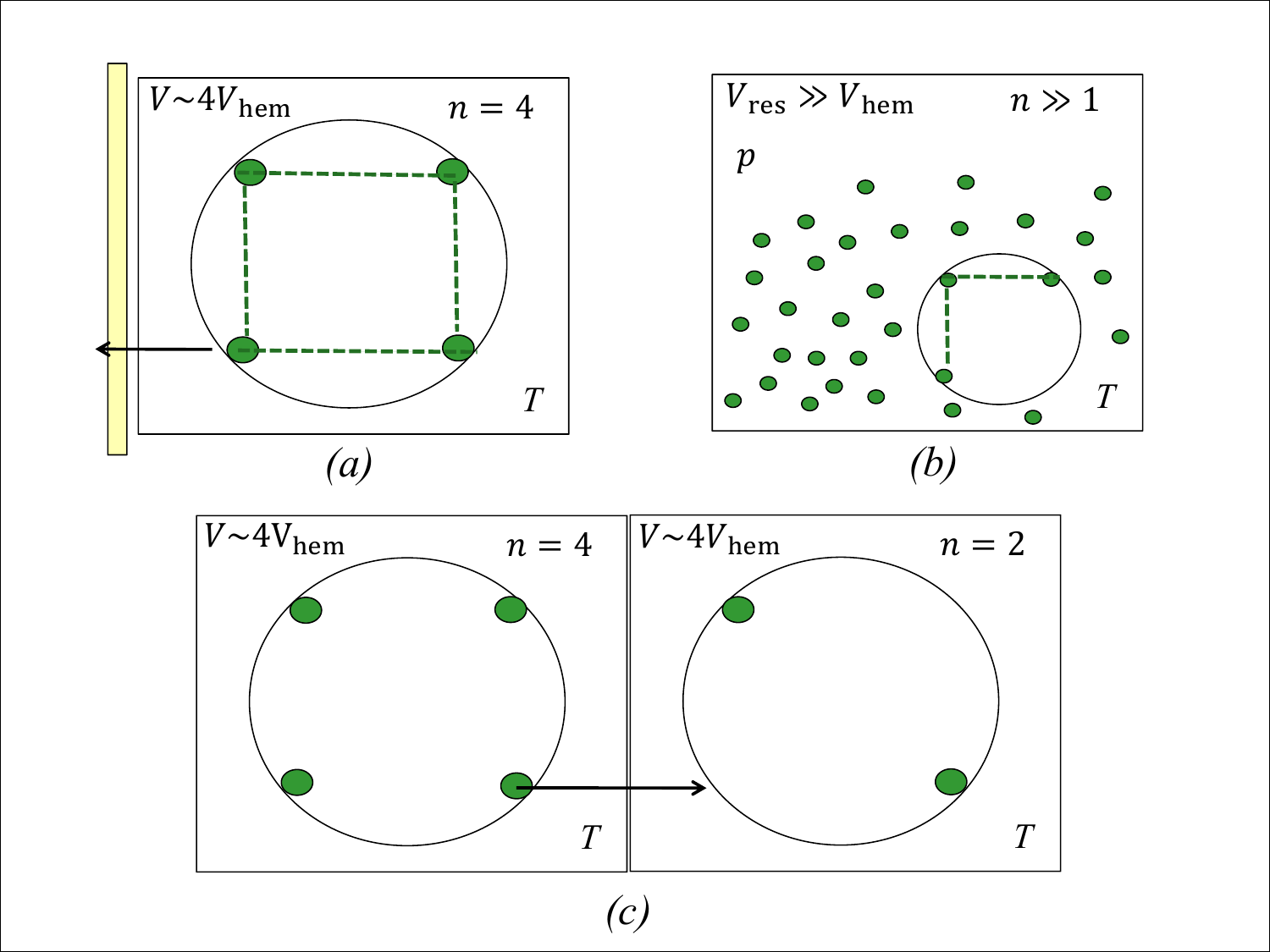}
    \caption{{\it (a)} The canonical ensemble for the hemoglobin (Hb) molecule (schematically a circle) with $n=4$ (fully bound) \O molecules immersed in a thermal bath (water molecules) at temperature $T$. The surrounding volume for (potentially) unbound \O molecules is $V$, which is $\sim 4$ times larger than the proper Hb volume $V_{\rm hem}$; see (\ref{loki1}) and Table~\ref{tab11}. Now $V$ can refer to a cube with linear size $6.93$ nm. 
    The canonical equilibrium is justified since each Hb macromolecule in RBC is surrounded by $\sim 1500$ water molecules; cf.~Table~\ref{tab11}. \\ 
    Dotted lines refer to Pauling's model, where bound \O molecules interact along the rectangular sides. The yellow bar denotes the RBC cellular membrane, via which the thermalized Hb can exchange particles with the external environment. The arrow indicates that a fully loaded Hb cannot attract more particles, but can release \O into blood plasma (or to other Hb); see (\ref{kau}). \\
    {\it (b)} The grand canonical ensemble: the Hb macromolecule with volume $V_{\rm hem}$ is immersed in an \O reservoir with partial pressure $p$. The reservoir volume holds $V_{\rm res}\gg V_{\rm hem}$, and $V_{\rm res}$ contains $n\gg 1$ number of \O molecules. The minimal realization of the particle reservoir with blood plasma is as follows. Given that the concentration of \O molecules in the alveolar blood plasma is $1.02\times 10^{15}\, {\rm molecules}/\mic^3$, $V_{\rm res}$ can refer to a cube with the linear size $0.1$ $\mic$, which will then contain $n\sim 100$ \O molecules. However, in RBC, such a $V_{\rm res}$ contains $\sim 10^3$ Hb macromolecules, and not a single Hb; cf.~Table~\ref{tab11}. Section \ref{grando} shows that the condition $n\gg 1$ per Hb molecule is needed for the grand canonical ensemble.
    \\
    {\it (c)} Two Hb macromolecules with (resp.) $n=4$ and $n=2$ in contact. The \O transport is governed by free energies; see (\ref{baran}). As expected, for a moderate cooperativity, the \O numbers tend to equalize; see the arrow. This conclusion does not hold for stronger cooperative interactions; see the discussion around (\ref{cascade}). }
    \label{fig-1}
\end{figure}

The standard description of the Hb-\O equilibrium starts with the grand canonical ensemble, where a single Hb molecule is immersed into a particle reservoir of \O molecules, described by a definite partial \O pressure \cite{dill,popel,pauling_pnas,eaton_models,mwc,changeux_review_50,koshland,ising}. 
The reservoir amounts to many unbound \O molecules dissolved in the RBC (red blood cell or erythrocyte)   cytoplasm or in blood plasma; see Fig.~\ref{fig-1}. Hence, the Hb-\O equilibrium can be monitored by the mean number of \O molecules bound to Hb \cite{pauling_pnas,eaton_models,mwc,changeux_review_50,koshland,ising}. 
The current understanding of Hb as a cooperative molecule is based on the grand canonical ensemble \cite{dill}. 

In the context of Hb and its basic function of \O transport, the grand canonical ensemble has the following drawback. Within the RBC, the concentration of plasmatic \O is at least 100 times lower than the concentration of \O bound to Hb \cite{popel}; see section \ref{cano}. This is the very reason why the whole machinery of \O transport via Hb exists (the \O dissolved in the blood is not sufficient for organisms). Hence, even in arterial blood (where Hb is almost fully ${\rm O_2}$-loaded), the majority of Hb molecules do not see any cytoplasmic \O molecule in their vicinity; each Hb sees only the 4 \O molecules attached to it. This is far from the standard setting of the grand canonical ensemble, where a single Hb is immersed into a reservoir of \O molecules; see Fig.~\ref{fig-1}. 

We resolve this problem by looking at a single Hb, which together with $n\leq 5$ oxygen molecules (bound or located at its vicinity) is in the canonical equilibrium with a fixed $n$ and temperature $T$; see Table~\ref{tab-ra} for the comparison of canonical {\it versus} grand canonical ensembles. The canonical ensemble demands an additional parameter: the volume $V$ around the Hb available for motion of unbound \O molecules. The choice of $V$ is not free, since Hb molecules are tightly packed in RBC. This demands $V$ to be larger but comparable to the proper Hb volume $V_{\rm hem}$; see Fig.~\ref{fig-1}. The grand canonical ensemble is recovered as a limiting case for $n\gg 1$, and $V=V_{\rm res}\gg V_{\rm hem}$, where only $n/V\propto p$ enters into pertinent quantities, with $p$ being the partial \O pressure, and $V_{\rm res}$ is the particle reservoir volume; see section \ref{grando}.

The main difference between the canonical ensemble and the grand canonical one is as follows. In the latter, the oxygen transport is controlled by the mean number of bound \O molecules, as a function of the partial pressure of the \O-reservoir. In the former, the absorption/release of \O from/to blood plasma 
is governed by differences of the canonical free energy as a function of $n$. The free energy mechanism puts additional  {thermodynamical} constraints on Hb models (such constraints are not effective for the grand canonical ensemble). It allows to reassess the role of cooperativity in \O binding by Hb. In addition, the canonical ensemble allows to describe \O transport in between of Hb molecules; this mechanism is absent for the grand canonical ensemble. 


Two major models of cooperative oxygen binding to hemoglobin -- Pauling's model \cite{pauling_pnas,eaton_models,ising} and MWC model \cite{mwc,changeux_review_50} -- are standardly studied within the grand canonical ensemble. These models implement different mechanisms of cooperativity: Pauling's model assumes interaction energy between the binding sites, while the MWC model introduces two different states for Hb \cite{pauling_pnas,eaton_models,mwc,changeux_review_50,koshland,ising}. Historically, Pauling's model was proposed earlier than the MWC model \cite{pauling_pnas,mwc}. The current consensus is that the MWC model and its generalizations are a better description of Hb-\O equilibrium and related features \cite{eaton_models,ahmed,bellelli}. However, Pauling's model is useful for explaining a possible cooperativity scenario \cite{dill}.

We show that these models can be formulated within the above canonical ensemble. The grand canonical ensemble is a particular case of this more general formulation; see section \ref{cano}. Reassessing Pauling's model and MWC model, 
we find that (for both models) the canonical ensemble consistently predicts lower values for the cooperativity parameters. Moreover, we demonstrate that within the canonical approach, higher values of cooperativity parameters will lead to various forms of instability, e.g., a cascade process of \O release in capillaries. The identification of these instabilities is a conceptual advantage of the canonical approach: within the grand canonical ensemble, the cooperativity parameters do not have such bounds. 
As a result, the instabilities provide a suitable test setup for checking the values of cooperativity parameters.  

\comment{
we find that there are important physical differences compared with the standard grand canonical situation. 
First of all, the cooperativity parameters are obtained from different quantities: within grand canonical ensemble, one looks at the mean number of bound particles. In the canonical situation, the relevant quantity is the free energy, since it governs the thermodynamics of diffusion. }

Another important difference between the ensembles is that the physical picture of the \O transport function of the Hb macromolecule within the canonical ensemble is machine-like: the \O is tightly bound to Hb and is released/accepted in discrete steps nearly without fluctuations. As compared to its mean, the standard deviation of the bound \O molecule number is suppressed. This contrasts the grand canonical ensemble, where Hb works as a fluctuating, statistical system with a standard deviation of the same order as the mean number of bound \O molecules.

Note that there are several pertinent physical situations where previous literature has detected important differences between canonical and grand canonical ensembles. Ref.~\cite{blumenfeld} examined such differences for chemical reactions in the low concentration regime; the approach of Ref.~\cite{blumenfeld} was criticized in Ref.~\cite{sokirko}. The differences are closely related to the validity of the mass action law in chemical kinetics \cite{mass_action_law}. Refs.~\cite{reiss,shegelski,kaplan} studied them in the context of quasi-free electron models in solid state physics. An important point of Refs.~\cite{reiss,shegelski,kaplan} is that the differences persist even for systems that are macroscopic under the operational meaning of this concept. 

The paper is organized as follows. Section \ref{cano} studies \O-Hb equilibrium within the canonic ensemble. 
Thermodynamics of diffusion is discussed in section \ref{frea}. Pauling's model is studied in section \ref{paul}. The obtained results are compared to those found from the grand canonical ensemble; see Appendix \ref{apo} for details. The MWC model is worked out in section \ref{monod}. We summarize in the last section, and underline directions for future research, which, in particular, should include non-equilibrium versions of the studied models.

\begin{table}[t]
\centering
\caption{Canonical {\it versus} grand canonical ensemble for oxygen-hemoglobin (\O-Hb) equilibrium. }
\begin{tabular}{|c|c|}
\hline
Canonical ensemble              & Grand canonical ensemble \\
\hline\hline
Temperature $T$ is fixed       & Temperature $T$ is fixed \\ 
\hline
Number of \O molecules $n$ around a single Hb and the volume $V$ 
& Partial pressure $p$ of \O reservoir is fixed; see (\ref{parto}) \\
available for unbound motion of \O molecules are fixed; see (\ref{h1}--\ref{h3}) & \\
\hline
\O transfer between Hb and plasma is governed 
& \O transfer between Hb and plasma
is governed \\
by the free energy difference (\ref{free})  
& by the number $\overline{n_b}(p)$ of bound \O molecules \\
& as a function of $p$; see (\ref{zhuk2}) \\
\hline
\O transfer between Hb molecules is governed 
& \O transfer between Hb molecules
is undefined \\
by the free energy difference (\ref{baran}) & \\
\hline
\end{tabular}
\label{tab-ra}
\end{table}

\begin{table}[t]
\centering
\caption{Quantities relevant for RBC (erythrocyte) and hemoglobin (Hb) \cite{data,richardson_pnas} under normal
conditions. }
\begin{tabular}{|c|c|l|}
\hline
Notation              & Value                                     & {Meaning}                   \\ \hline
$n_{\rm O_2}$(blood)  & $1.\times 10^{5}$ molecules/$\mic^3$     & Concentration of ${\rm O_2}$ in alveolar blood plasma at 100 torr           \\ \hline
$n_{\rm O_2}$(water)  & $1.5\times 10^{5}$ molecules/$\mic^3$      & Concentration of ${\rm O_2}$ in fresh water at 100 torr           \\ \hline
$n_{\rm O_2}$(air)    & $5\times 10^{6}$ molecules/$\mic^3$       & Concentration of ${\rm O_2}$ in air at 100 torr           \\ \hline
$n_{\rm H_2O}$         & $2.4\times 10^{10}$ molecules/$\mic^3$   & Concentration of ${\rm H_2O}$ in RBC                                 \\ \hline
$n_{\rm Hb}$           & $3 \times 10^{6}$ molecules/$\mic^3$     & Concentration of ${\rm Hb}$ in RBC                                 \\ \hline
$v_{\rm Hb-O_2}$       & $50-100 \times 10^{-12}$ $\mic^3$        & The volume of the ${\rm Hb}$ heme pocket for ${\rm O_2}$ molecule    \\ \hline
$r_{\rm Hb}$         & $2.5\times 10^{-3}$ $\mic$                 & Radius of  ${\rm Hb}$ molecule approximated  as a sphere  \\ \hline
$n_{\rm RBC}$         & $5\times 10^{9}$ cells/${\rm cm}^3$            & Concentration of RBC   \\ \hline
$N_{\rm Hb|RBC}$         & $3\times 10^{8}$                       & Number of Hb in RBC   \\ 
\hline
$V_{\rm RBC}$         & $90$ $\mic^3$                             & Volume of RBC   \\ 
\hline\hline
$D_{\rm O_2|H_2O}$       & $2.5\times 10^{3}$ $\mic^2$/s          & Diffusion constant of ${\rm O_2}$ in ${\rm H_2O}$ ($T=310$ K) \\ \hline
$D_{\rm O_2|RBC+memb}$       & $7\times 10$ $\mic^2$/s      & Diffusion constant of ${\rm O_2}$ in RBC (+ membrane permeability) \\ \hline
$D_{\rm Hb|RBC}$       & $17.5$ $\mic^2$/s      & Diffusion constant of Hb in RBC including crowding effects \\ \hline
\end{tabular}
\label{tab11}
\end{table}

\section{Model for a hemoglobin-oxygen equilibrium}
\label{cano}

\subsection{Criticism of the standard approach }
\label{svede}

The standard approach amounts to describing the Hb-\O equilibrium via the grand canonical approach, where a single Hb molecule is immersed in the reservoir of plasmatic (unbound) \O molecules; see Fig.~\ref{fig-1}. A necessary condition for this is that the concentration $n_{\rm O_2}$ of plasmatic \O is much larger than the concentration $n_{\rm Hb}$ of Hb in RBC. (We neglect the equilibrium partial pressure differences between blood plasma and RBC cytoplasm.) This condition does not hold in RBC: $n_{\rm Hb}/n_{\rm O_2}\simeq 30$; see Table~\ref{tab11}. This estimate can be generalized beyond RBC: we can ask for Hb concentration $n_{\rm RBC}N_{\rm Hb|RBC}$ in the blood, which is the concentration $n_{\rm RBC}$ of RBC in the blood times the number $N_{\rm Hb|RBC}$ of Hb's in a single RBC. Even this number is still larger than $n_{\rm O_2}$: $n_{\rm RBC}N_{\rm Hb|RBC}/n_{\rm O_2}\simeq 15$; cf.~Table~\ref{tab11}. Further details on the applicability of the grand canonical ensemble to the Hb-\O equilibrium are discussed in section \ref{grando} and Appendix \ref{apo_clarifo}.

\subsection{Canonical ensemble}

We consider a single Hb molecule with overall volume $V_{\rm hem}$ and 4 traps. Their volumes are specified below. The free volume (where unbound \O molecules move) is $V$; see Fig.~\ref{fig-1}. There are $n$ \O molecules, and no two particles can enter the same trap. Unbound \O molecules do not interact, while bound molecules do (cooperativity). This interaction is specified below. The equilibrium coordinate density of $n$ particles reads \cite{landau5}:

\comment{but the number of already bound molecules alters the energy of binding and its volume: i.e. if $k-1$ molecules are already bound, the next bound molecule sees the binding energy $\varepsilon(k)$ and the binding volume $v(k)$. This 
}

\BEA
\label{h0}
P(\r_1,...,\r_n)=\frac{1}{{\cal Z}} e^{-\beta U(\r_1,..,\r_n)}, \qquad
{\cal Z}=\frac{1}{n!}\int\prod_{k=1}^n{\rm d}\r_k e^{-\beta U(\r_1,..,\r_n)},
\EEA
where ${\cal Z}$ is the coordinate part of the canonical partition function, $\beta=1/T$ is the inverse temperature ($k_{\rm B}=1$), and where $U(\r_1,..,\r_n)$ is the potential energy, which is invariant with respect to all permutations of $(\r_1,..,\r_n)$. We note:
\BEA
\label{urs}
\int {\rm d}\r_k=\sum_{i=0}^4\int_{\Omega_i} {\rm d}\r_k,
\EEA
where $\Omega_0$ with ${\rm volume}[\Omega_0]=V$ is the space domain of the unbound motion, and where $\Omega_i$ includes the binding domain $i$ and its vicinity ($i=1,...,4$). In
$\int_{\Omega_1} {\rm d}\r_k$ the integration is restricted to $\Omega_1$. Putting (\ref{urs}) into (\ref{h0}), we get
\BEA
\label{h7}
{\cal Z}=\frac{1}{n!}\sum_{k_1,...,k_n=0}^4\zeta(k_1,...,k_n),
\EEA
where (say) the index $k_1=0,...,4$ in (\ref{h7}) counts the case when the first particle is unbound ($k_1=0$), is bound to the first binding volume ($k_1=1$), {\it etc}. We assume $U(\r_1,..,\r_n)$ is such that $\zeta(k_1,...,k_n)=0$ if among $(k_1,...,k_n)$ there are at least two coinciding non-zero indices, e.g., $k_1=k_2>0$. This corresponds to two particles being in a single well, which is prohibited. We assume that unbound particles move in volume $V$ and do not interact with each other. Bound particles interact with each other. The interaction mechanism depends on the number of bound particles, but (naturally) does not depend on their label. The mechanism is illustrated as follows for $n=4$:
\BEA
\zeta(0,0,0,1)=\int_{\Omega_0} {\rm d}\r_1\int_{\Omega_0} {\rm d}\r_2\int_{\Omega_0} {\rm d}\r_3
\int_{\Omega_1} {\rm d}\r_4\, e^{-\beta U(\r_1,\r_2,\r_3,\r_4)}=
V^3 v(1) e^{-\beta \varepsilon(1) },
\EEA
where only the fourth particle is bound in $\Omega_1$, $v(1)$ is the corresponding binding volume, and $\varepsilon(1)$ is the energy. Neither $v(1)$ nor $\varepsilon(1)$ are dependent on which molecule got bound and where it was bound.
Likewise,
\BEA
\zeta(0,0,1,2)=\int_{\Omega_0} {\rm d}\r_1\int_{\Omega_0} {\rm d}\r_2\int_{\Omega_1} {\rm d}\r_3
\int_{\Omega_2} {\rm d}\r_4 \, e^{-\beta U(\r_1,\r_2,\r_3,\r_4)}
=V^2v(1)v(2)e^{-\beta \varepsilon(1)-\beta \varepsilon(2)},
\EEA
where the second bound molecule sees a different binding volume $v(2)$ and energy $\varepsilon(2)$. Again, these parameters do not depend on the number of the particle and the number of the trap. The differences $v(2)-v(1)$ and $\varepsilon(2)-\varepsilon(1)$ allow for cooperativity within the induced fit approach. 
(The MWC model employs a different mechanism of cooperativity, but still fits within the present model; see section \ref{monod}.)

Hence, if $(k_1,...,k_n)$ contains $\ell$ ($1\leq \ell\leq 4$) non-zero indices (i.e., $\ell$ bound particles and $n-\ell$ unbound ones), we get
\BEA
\label{ton}
&&\zeta(k_1,...,k_n)=V^{n-\ell}\prod_{i=1}^\ell v(i)e^{-\beta \varepsilon(i)}
=V^{n-\ell}\prod_{i=1}^\ell e^{-\beta f(i)},\\
\label{h3}
&&f(i)=\varepsilon(i)-T\ln [v(i)/V],
\EEA
where $f(i)$ is the (effective) free energy. 
The number of different factors $\zeta(k_1,...,k_n)$ that satisfy (\ref{ton}) equals 
$\ell ! \binom{n}{\ell}\binom{4}{\ell} $, where $\binom{n}{\ell}\equiv\frac{n!}{\ell!(n-\ell)!}$ 
comes from the number of $\ell$ places within $(k_1,...,k_n)$,
$\ell !$ is responsible for permutation of non-zero indices in $(k_1,...,k_n)$, and 
$\binom{4}{\ell} $ refers to the selection of $\ell$ binding volumes out of $4$. 
We have from (\ref{h7}, \ref{ton}):
\BEA
{\cal Z}=\frac{1}{n!}V^n+\frac{1}{n!}\sum_{\ell=1}^4 \ell !\binom{n}{\ell} \binom{4}{\ell}
V^{n-\ell}\prod_{i=1}^\ell v(i)e^{-\beta \varepsilon(i)}\equiv \frac{1}{n!}V^n Z.
\label{zzz}
\EEA
The probability $p(k)$ that $k=0,...,4$ molecules are bound is found from (\ref{zzz}, \ref{h3}):
\BEA
\label{h1}
&&p(0)=\frac{1}{Z},\\
\label{h2}
&&p(k)=\frac{1}{Z}\,\binom{4}{k} \, n(n-1)...(n-k+1)
\prod_{i=1}^k e^{-\beta f(i)}, \quad k=1,...,4. 
\EEA
The mean number of bound particles reads from (\ref{h2}):
\BEA
\label{hmean}
\overline{n_b}
=\sum_{i=1}^4 ip(i).
\EEA
\comment{
In (\ref{h3}), we took $\varepsilon(0) = 0$, hence $\varepsilon(k > 0)<0$. But $-T\ln [v(i)/V]>0$ in (\ref{h3}), which is the usual competition between energy and entropy. 
}

\comment{
Let us write down $Z$ in (\ref{h2}) explicitly:
\BEA
&&Z=1+4n e^{-\beta f(1)}+6n(n-1)e^{-\beta f(1)-\beta f(2)}+4n(n-1)(n-2)e^{-\beta f(1)-\beta f(2)-\beta f(3)}\nonumber\\
&&+n(n-1)(n-2)(n-3)e^{-\beta f(1)-\beta f(2)-\beta f(3)-\beta f(4)}.
\label{z}
\EEA
}

We focused on the coordinate distribution (\ref{h0}) and the corresponding partition function (\ref{zzz}). This is sufficient for our purposes. Appendix \ref{apo_full} recalls the full partition function
of the model, also including the momentum part.

\subsection{Parameters of the canonical distribution}

\comment{
The model requires that for a finite $n$ (e.g., $n\leq 5$), the volume $V$ of free motion is not too large. Indeed, estimating $v\simeq v(i)$ via the heme pocket size $v_{\rm Hb-O_2}$ [see Table~\ref{tab11}], we see that $V\gg v$ will eventually lead to in (\ref{h3}) to $f(i)>0$, i.e. it will prevent \O binding.} 

The very structure of RBC, together with our intention of describing a single Hb molecule, estimates the volume $V$ around a Hb molecule as [see Table~\ref{tab11}]:
\BEA
\label{loki1}
V=\frac{1}{ n_{\rm Hb}} -\frac{4\pi r_{\rm Hb}^3 }{3}
\simeq 4.1\times\frac{4\pi r_{\rm Hb}^3 }{3}\simeq 2.68\times 10^{-7}\, \mu{\rm m}^3,
\EEA
$n_{\rm Hb}$ is the Hb concentration in the RBC, and $r_{\rm Hb}$ is the Hb radius (the Hb is approximated as a sphere), and $\frac{4\pi r_{\rm Hb}^3 }{3}$ is the Hb volume.
 
\comment{
To estimate $V$, recall that the Hb macromolecule is approximated as a sphere with radius $r_{\rm Hb}=2.5$ nm \cite{popel}. Since Hbs are densely packed in the erythrocyte, we can take $V=6^3\, {\rm nm}^3$. Recalling that the concentration of ${\rm O}_2$ in blood plasma is $1.02\times 10^{17}\,{\rm molecules}\cdot {\rm cm}^{-3}$, we see again that the concentration of ${\rm O}_2$ carried out by a fully loaded Hb is 185 larger than in the blood plasma.  
}

Let us discuss the characteristic relaxation times of the model. It roughly amounts to relaxation of the \O molecule. The latter relaxation process consists of two steps. First, the molecule diffuses in water to reach the Hb. This time estimates as $\tau_1\sim (2\,{\rm nm})^2/D_{\rm O_2|H_2O}\sim 1.6\times 10^{-9}$ s, where $D_{\rm O_2|H_2O}$ is the diffusion coefficient of \O in ${\rm H_2O}$; see Table~\ref{tab11} 
Second, \O reaches from the surface of the heme pocket to its interior. This takes longer time $\tau_2\sim 10^{-7}$ s \cite{relaxation1}.

Note that applying the concept of diffusion coefficient requires faster local equilibration, which is satisfied here. The relaxation time $\frac{1}{\gamma}$ of an \O molecule in the water thermal bath can be estimated from the friction term in the Langevin equation as $m_{{\rm O}_2}\gamma\simeq r_{{\rm O}_2}\eta_{\rm wat}$, where $m_{{\rm O}_2}=5.3\times 10^{-23}$ g and $r_{{\rm O}_2}=0.152$ nm are (resp.) the mass and radius of ${\rm O}_2$, and $\eta_{\rm wat}=0.01\, {\rm g}\cdot{\rm cm}\cdot {\rm s}^{-1}$ is the water viscosity. The rationale behind this standard estimate was recently reviewed in Ref.~\cite{ma_pre}, which also discussed that since $m_{{\rm O}_2}\sim m_{{\rm H}_2{\rm O}}$, the friction in the Langevin equation is not memoryless. This fact does not alter the order of magnitude estimate of the relaxation time presented here \cite{ma_pre}. This produces for the fast local relaxation time $\frac{1}{\gamma}\sim 10^{-13}$ s.

\subsection{Grand canonical ensemble}
\label{grando}

The transition to the grand canonical ensemble is carried out as follows. In (\ref{zzz}), we assume that $n\gg 1$, and 
approximate uniformly $\ell ! \binom{n}{\ell}\simeq n^\ell$. Hence $p(k)$ in (\ref{h2}) contain $n$ and $V$ only in combination $n/V$ (concentration). Hence, from the viewpoint of (\ref{h2}), the Hb molecule is immersed into a particle reservoir with concentration $n/V$. This is the way the grand canonical ensemble is applied in practice, because the particle reservoir can now be the blood plasma; see Table~\ref{tab-ra}. 
Now $V=V_{\rm res}$ is the reservoir volume. It should also be large, but both $n$ and $V$ can be left unspecified, since only $n/V$ enters into $p(k)$. 
Using the ideal gas law, we can express $n/V$ via the partial \O
pressure $p$ in air, which is controlled externally:
\BEA
\label{parto}
p=\frac{n}{V}\frac{T}{H_G}, 
\EEA
where $H_G$ is Henry's constant for air-blood plasma equilibrium. We get from (\ref{zzz}, \ref{hmean}) for the partition function and the mean number of bound \O molecules in the grand canonical ensemble:
\BEA
{\cal Z}_{\rm GC}=1+
\sum_{\ell=1}^4 p^\ell \binom{4}{\ell}
\prod_{i=1}^\ell \frac{H_G}{T}v(i)e^{-\beta \varepsilon(i)},
\label{zhuk},\\
\overline{n_b}=\frac{1}{Z_{\rm GC}}\Big[ 
\sum_{\ell=1}^4\ell p^\ell\binom{4}{\ell}
\prod_{i=1}^\ell \frac{H_G}{T}v(i)e^{-\beta \varepsilon(i)}
\Big].
\label{zhuk2}
\EEA
Eq.~(\ref{hmean}) is similar to Adair's equation \cite{adair,dill}.
For the grand-canonical situation, $\overline{n_b}$ in (\ref{zhuk2}) is the basic diagnostic quantity, which, in particular, allows one to decide on cooperativity. For the canonical situation, $\overline{n_b}$ does not have the same diagnostic meaning. This role is played by the free energy, which is discussed below. 

Applicability limits of canonical and grand canonical ensembles are indeed different; cf.~ Table~\ref{tab-ra}. One can ask whether the above results on the grand canonical ensemble can be derived more generally by including into an equilibrium description all \O molecules and Hb molecules of the RBC. Appendix \ref{apo_clarifo} answers this question negatively. 

We mention here another salient feature of the grand canonical ensemble: assume an isolated Hb in alveolar blood. To immerse this Hb into an \O particle reservoir, we can assume that the characteristic size of this reservoir is $\simeq 0.1\mic$. Then Hb will be surrounded by $\simeq 100$ \O molecules; cf.~Table~\ref{tab11}. Surrounded means that any reservoir particle can reach the Hb by diffusion. Noting the value of 
the diffusion constant $D_{\rm O_2|H_2O}$ from Table~\ref{tab11}, we see that the time is $\simeq 0.4$ s. For shorter processes, the grand canonical ensemble does not apply to the Hb. 

\section{Free-energy and thermodynamics of diffusion}
\label{frea}

\subsection{\O transport with the outer environment }

\subsubsection{Free energy differences}

The free energy of the Hb with $n$ \O molecules reads from (\ref{zzz}, \ref{h1}, \ref{h2}):
\BEA
 F[n]=-T\ln{\cal Z}
 =-T(n\ln[V]+\ln Z)+T\ln[n!].
\label{brut}
\EEA
Consider a Hb molecule in contact with the RBC membrane; see Fig.~\ref{fig-1}. We can employ (\ref{brut}) for determining the \O transport between the Hb and the \O dissolved in the blood plasma. To this end, we assume that $\hat N$ \O molecules dissolved in the blood plasma make an ideal dilute solution in volume $\hat V$ and temperature $T$. The corresponding free energy, and the total free energy read (resp.)
\BEA
&& \hat F[\hat N]=-T \hat N\ln \hat V+T\ln[\hat N!],\\
&& { F}_{\rm tot}[n,\hat N]=\hat F[\hat N]+ F[n]. 
\label{freedom}
\EEA
We did not include in (\ref{brut}--\ref{freedom}) the partition function coming from the kinetic energy of \O molecules [see Appendix \ref{apo_full}], because it is $\propto (\hat N+n)$, and hence it will cancel out from $\Delta {F}_{\rm tot}$, as seen below. 

\comment{
This statement implies time-scale separation: first the two systems equilibrate separately, and then the diffusion tends to decrease (ot at least not to increase) the total free energy. The time-scale separation is behind the standard description of diffusion \cite{landau6}. Let us argue that it is present here. 

Now recalling the diffusion constant $D_{ {\rm O}_2| {\rm blood}  }=2.1\times 10^{-5} {\rm cm}^2\cdot {\rm s}^{-1}$ of ${\rm O}_2$ in the blood plasma under normal conditions, we note that within $100\frac{1}{\gamma}$, ${\rm O}_2$ will move $\sim 1$ nm due to diffusion. \textcolor{red}{Please look up Richardson's paper. There is an effective diffusion constant equal to 70~$\mu\text{m}^2$/s. Here is the link: \url{https://doi.org/10.1073/pnas.1916641117}.}
}

The \O redistribution across the membrane tends to minimize the free energy (\ref{freedom}).
If a single \O molecule moves from the Hb to the plasma during a quasi-equilibrium process, the free energy change reads
\BEA
&&\Delta {F}_{\rm tot}\equiv { F}_{\rm tot}[n,\hat N]-{ F}_{\rm tot}[n-1,\hat N+1]
=T\Big(g[n]-\ln\frac{V(\hat N+1)}{ \hat V }\Big),
\label{free}
\\
&&g[n]\equiv \ln\frac{nZ[n-1]}{Z[n]}.
\label{mree}
\EEA
$\Delta {F}_{\rm tot}<0$ in (\ref{free}) means that \O absorption by the Hb is more likely statistically. 
This is most clearly seen within a non-equilibrium description, where the free-energy difference determines the transition rate from one state to another \cite{dill}. In the same sense, $\Delta {F}_{\rm tot}>0$ means \O release. Note that when the reasoning of (\ref{brut}--\ref{freedom}) is applied to two ideal gases (dilute solutions), we get - consistently with the most usual understanding of diffusion - that the particle has statistical tendency to go from a higher concentration to a lower one.

\comment{
For the ${\rm O}_2$ concentration (under normal conditions) in air, fresh water and blood plasma we have respectively: $5\times 10^{18}\,{\rm molecules/cm}^3$, $1.6\times 10^{17}\,{\rm molecules/cm}^3$, and $7.2\times 10^{16}\,{\rm molecules/cm}^3$. 
For a single Hb molecule, $V$ naturally has the same order of magnitude as the volume of the Hb itself: $\frac{4\pi}{3}(2.5\times 10^{-7}\, {\rm cm})^3$. 
The precise value of $V$ is, of course, chosen conventionally, but increasing it 5 times (which is unacceptable) will change the results below by a relatively small additive factor $\ln 5=1.61$.
}

\subsubsection{Bounds for free energy differences in \O transport}

We want to find a realistic estimate for the parameter $(\hat N+1)/\hat V$ in (\ref{free}).
Given the volume $V_{\rm RBC}$ of RBC (and approximating it as a sphere) the surface of RBC is estimated as 
$S_{\rm RBC}=(36\pi)^{1/3}V_{\rm RBC}^{2/3}$; see Table~\ref{tab11}. In the present model, we surrounded each Hb molecule by the cytoplasm volume $V$ so that each Hb+$V$ amounts to a cube with side 6.93 nm; cf.~Fig.~\ref{fig-1}. Thus, the number of Hb molecules located next to the RBC surface is estimated as [cf.~Table~\ref{tab11}]:
\BEA
\label{loki2}
N_{\rm Hb|surf.\,RBC}\simeq S_{\rm RBC}/(6.93\, {\rm nm})^2\simeq 2\times 10^6.
\EEA
In the blood plasma, the number of \O molecules per a RBC can be estimated with the same idea as for (\ref{loki1}).
For the plasma volume per RBC we get $n_{\rm RBC}^{-1}-V_{\rm RBC}\simeq 1.1\times 10^{-11}{\rm cm}^3$, where $n_{\rm RBC}$ is the RBC concentration; see Table~\ref{tab11}. Thus the number of plasmatic \O molecules per RBC reads:
\BEA
\label{loki3}
N_{\rm O_2|RBC}\simeq(n_{\rm RBC}^{-1}-V_{\rm RBC}) n_{\rm O_2}\simeq 1.1\times 10^7,
\EEA
where $n_{\rm O_2} $ is the \O concentration in alveolar blood; see Table~\ref{tab11}.
Comparing (\ref{loki2}) and (\ref{loki3}), we see that there are $\simeq 5$ \O molecules per one Hb molecule, which is located at the RBC surface. Now we can estimate $\frac{\hat N+1}{ \hat V }$ in (\ref{free}, \ref{mree}) for alveolar blood. We choose for $\hat V$ the smallest volume which is about to contain $\zeta=5$ \O molecules, i.e., $\hat N=\zeta-1$ and $\hat V=\zeta/n_{\rm O_2}$. Together with (\ref{loki1}), this produces for the alveolar blood plasma
\BEA
\ln\Big(\frac{V(\hat N+1)}{ \hat V }\Big)
\simeq 
\m=-3.62. 
\label{bela}
\EEA
Eq.~(\ref{bela}) does not depend on the concrete value of $\zeta$, provided that $\zeta\geq 1$.

\begin{table}[t]
\centering
\caption{Relationship between the partial oxygen pressure $p$ in blood plasma [cf.~(\ref{parto})], Hb saturation ${\rm SaO}_2$, and clinical interpretation of oxygenation. At 100 torr (lungs) the \O concentration in blood plasma is $10^{17}\, {\rm molecules/cm}^3$; see Table~\ref{tab11}. For other values of $p$, the concentration is recalculated linearly.\\
The above data is taken from Ref.~\cite{data}. It is measured at standard physiological conditions: pH$\simeq 7.4$ in blood plasma (pH$\simeq 7.24$ in RBC); the ratio of concentrations [2,3‑BPG]/[Hb]$\simeq 0.9$ ([2,3‑BPG] = 4.65 mM); $T=37\,^{\circ}\mathrm{C}$\,. The data for $p\leq 20$ torr is consistent with interpolations of Severinghaus and Hill equations \cite{severinghaus}. 
}
\begin{tabular}{|c|c|l|}
\hline
$p$ (torr) & ${\rm SaO}_2$ (\%)  & \textbf{Interpretation}       \\ \hline
80                   & 95    & Almost full                                 \\ \hline
60                   & 90    & Beginning of \O release                 \\ \hline
40                   & 75    & Resting tissue \O delivery              \\ \hline
30                   & 55    & Active tissue (e.g. muscle during exercise) \\ \hline
20                   & 35    & Very active tissue (vigorous muscle activity)  \\ \hline
10                   & 15    & Severe hypoxia                 \\ \hline
5                   &  5    &  Critical hypoxia               \\ \hline
\end{tabular}
\label{tab1}
\end{table}
The minimal condition for ensuring that the Hb is fully loaded in lungs is deduced from (\ref{free}--\ref{bela}):
\BEA
\label{c1}
g[n]<\m,\qquad 4\geq n\geq 1.
\EEA
Table~\ref{tab1} presents a standard data about Hb saturation. These data is obtained from macroscopic measurements and suggests the following (to some extent conventional, but reasonable) conditions to be applied to a single Hb 
\BEA
\label{test4}
& \m-\ln[10/4]<&g[4]<\m-\ln[10/8], \\
\label{test3}
& \m-\ln[10/3]<&g[3]<\m-\ln[10/4],\\
\label{test2}
& \m-\ln[10/2]<&g[2]<\m-\ln[10/3],\\
\label{test1}
& \m-\ln[100/7.5]< & g[1]< \m-\ln[10/2],
\EEA
where (\ref{test4}) means that the first ${\rm O}_2$ molecules is released to the outer environment when 
$p$ in blood plasma decreases from 80 torr to 40 torr. According to (\ref{test3}) the second 
${\rm O}_2$ is released when $p$ in blood plasma changes from 40 to 30 torr. The third \O
molecules is released when $p$ changes between 30 and 20 torr; see (\ref{test2}). 
Finally, the last \O molecule is released when $p$ changes between 20 and 7.5 torr, as (\ref{test2}) shows. We emphasize again that the bounds presented in (\ref{test4}--\ref{test1}) are somewhat conventional, because we had to decide how to fit 7 (approximate) numbers to 4 released \O molecules. At the present stage of our understanding, such conventionalities are unavoidable. 

Note from (\ref{zzz}, \ref{mree}) that for $n\geq 5$, every term in $\frac{1}{n}Z[n]$ is smaller than the corresponding term in $Z[n-1]$; hence
\BEA
g[n]\geq 0, ~~{\rm for}~~ n\geq 5.
\label{kau}
\EEA
For $n<5$, $Z[n-1]$ contains fewer terms than $\frac{1}{n}Z[n]$, and $g[n]>0$ for $n<5$ becomes possible. The message of (\ref{kau}) is that even in alveols, the filled Hb will not attract more \O molecules into its vicinity. Indeed, such an attraction would require $g[5]<\m$, which is impossible due to (\ref{kau}). This conclusion is valid also for a two-molecule \O transfer from/to the blood; see Appendix \ref{2molecules}.    

\subsection{\O transport between two Hb molecules }
\label{tramspo}

We can apply (\ref{brut}) and the concept of the free energy decrease 
to the \O exchange between two Hb's with different values of $n$; e.g. denote by $(n+1;n-1)$ the pair with (resp.) $n+1$ and $n-1$ molecules. We get [cf.(\ref{free})]:
\BEA
\label{baran}
h[n]\equiv \frac{1}{T}(2F[n]-F[n+1]-F[n-1])=\ln\frac{nZ[n+1]Z[n-1]}{Z^2[n](n+1)}.
\EEA
Now $h[n]<0$ means stability of $(n;n)$ state against the spontaneous emergence of the gradient of $n$; i.e.
\BEA
h[n] <0 \iff (n+1;n-1)\to (n;n).
\label{baran0}
\EEA
The sign of $h[n]$ in (\ref{baran0}) is not fixed thermodynamically, i.e., the spontaneous emergence of the gradient is possible. 

The reasoning based on free energy determines statistical trends only. It is silent on characteristic {\it times} of \O transfer. Determining these times demands dynamic models, which we shall develop in the future based on the present static approach.

Note that the oxygen transport between two Hb molecules is not defined within the grand canonical ensemble.

\section{Different models of cooperativity}

The general framework developed above for the hemoglobin-oxygen equilibrium requires accounting for the cooperativity of oxygen binding to the Hb molecule. We will address below the implementation of cooperativity based on two widely known models, proposed (resp.) by Pauling \cite{pauling_pnas} and Monod, Wyman, and Changeux \cite{mwc}; see \cite{dill}.

\subsection{Pauling's model}
\label{paul}

This simplest, two-parametric model of Hb was proposed by Pauling \cite{pauling_pnas}. 
Later on, it was generalized in Ref.~\cite{koshland} within the induced fit (or sequential) approach \cite{eaton_models}. A seemingly different approach to hemoglobin was proposed, which is based on the one-dimensional Ising model \cite{ising}. However, this approach can be reduced to Pauling's model \cite{ising}.
We focus on Pauling's model, since this is the simplest and clearest example of the induced fit (or sequential) approach. 

The model amounts to introducing into (\ref{zzz}) a notation 
$K=e^{-\beta f(1)}$ and a new parameter $\alpha$:
\cite{pauling_pnas} 
\BEA
\label{pau1}
\frac{  e^{-\beta f(2)}}{   e^{-\beta f(1)}  }=\frac{2\alpha+1}{3},\qquad
\prod_{i=2}^3\frac{ e^{-\beta f(i)}}{   e^{-\beta f(1)}  }=\alpha^2, \qquad 
\prod_{i=2}^4\frac{  e^{-\beta f(i)}}{   e^{-\beta f(1)}  }=\alpha^4,
\EEA
where $T\ln\alpha$ is the additional free energy, stabilizing two interacting ${\rm HbO_2}$ groups \cite{pauling_pnas}. Now $Z[n]$ from (\ref{zzz}) reads
\BEA
Z[n]&=&1+4nK+ (4\alpha+2)n(n-1) K^2+4n(n-1)(n-2)\alpha^2 K^3\nonumber\\
\label{pau2}
&&+n(n-1)(n-2)(n-3)\alpha^4 K^4,\\
\label{cosmo}
&& p(0)=\frac{1}{Z},\quad p(1)=\frac{4nk}{Z},...
\EEA
The logics of introducing the cooperativity parameter $\alpha$ into (\ref{pau2}) follows Pauling's idea that the cooperativity in oxygen molecules binding is present only along the dotted lines of the square in Fig.~\ref{fig-1}$(a)$, but not along its diagonals \cite{pauling_pnas}. If only one \O molecule is bound, then no cooperativity is present; hence the factor $4K$ in (\ref{pau2}). If all molecules all present then there is full cooperativity along four sides os the square; hence $\alpha^4 K^4$ in (\ref{pau2}). Next, there are four arrangements of the three molecules, and the cooperativity goes along two sides of the square; hence, the factor $4\alpha^2 K^3$. Finally, $(4\alpha+2) K^2$ in (\ref{pau2}) is explained by noting that there are 6 possible arrangements of two \O molecules on the square. Still, two of them do not lead to cooperativity, since the molecules are arranged along the diagonals of the square. Note that cooperativity is described by $\alpha\geq 1$; no cooperativity means $\alpha=1$. The logic of introducing $\alpha$ remains thermodynamic and phenomenological. Mesoscopic features that enable the Hb cooperativity are discussed in \cite{hopfield,sasun}.

\begin{figure}[t]
    \centering
    \includegraphics[width=0.48\linewidth]{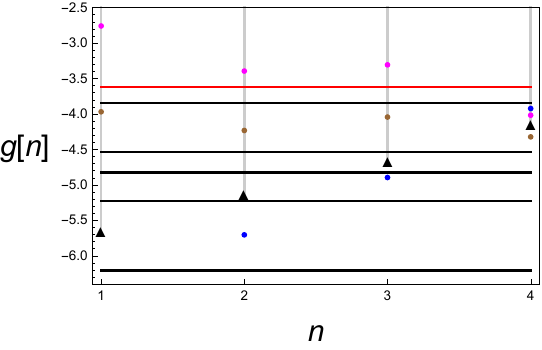}
    \caption{Free energy levels: $g[n]$ versus $n$ for Pauling's model; see (\ref{mree}, \ref{pau2}). Horizontal red line: $\m=-3.62$; cf.~(\ref{bela}). Four upper horizontal black lines refer (resp.) to the right-hand sides of (\ref{test4}--\ref{test1}). The lowest horizontal black line denotes $\m-\ln[100/7.5]$; see (\ref{test1}). 
    Black triangles, magenta points, brown points, and blue points are the values of $g[n]$ for (resp.) ($K=71.33$, $\alpha=1.88$), ($K=3.7$, $\alpha=7.83$), ($K=13$, $\alpha=4.83$), and ($K=200$, $\alpha=1$); cf.~(\ref{vard1}, \ref{bard2}). The latter case with $\alpha=1$ means no cooperativity. 
    Black triangles refer to the best fit (\ref{vard1}). They hold the inequalities in (\ref{test4}--\ref{test1}).  }
    \label{fig-3}
\end{figure}

The parameters $K$ and $\alpha$ of Paulings model are found by calculating $g[n]$ from (\ref{mree}, \ref{pau2}) and using (\ref{test4}--\ref{test1}) for fitting the values $\{g[n]\}_{n=1}^4$. For definiteness, $g[n]$ (for $n=4,3,2$) will be fit to the middle point of the corresponding interval in (\ref{test4}--\ref{test1}); e.g. $g[4]$ is fit to $\nu_0-\frac{1}{2}\ln[100/32]$. The best fit is obtained numerically under the least square criterion. 
It reads (recall that $\m=-3.62$):
\BEA
K=71.33, \quad \alpha=1.88.
\label{vard1}
\EEA
The obtained values of $g[n]$ are shown in Fig.~\ref{fig-3} as black triangles. It is seen that inequalities (\ref{test4}--\ref{test1}) are satisfied.

According to (\ref{vard1}), the cooperativity factor $\alpha$ is smaller than the prediction $\alpha=6.32$ found for the grand canonical implementation of Pauling's model, which is fit to the data of Table~\ref{tab1}; see around (\ref{bard2}) for more details. Recall that the value of $\alpha$ does not depend on the ensemble used; therefore, comparing them is legitimate. Pauling found even a larger value $\alpha=12.$ for the grand canonical model \cite{pauling_pnas}, but he employed the data by Ref.~\cite{ferry}. We shall see below that the difference between $\alpha=6.32$ and $\alpha=1.88$ is indeed physically substantial. 

Note that for different values of $\m$, only the fitted value of $K$ changes, while $\alpha$ stays nearly intact. For example, for $\m=-2.62$ and $\m=-5.1$ we get (resp.) $K=26.13$ and $K=313.98$ with the same $\alpha=1.88$.  

\begin{figure}[t]
    \centering
    \includegraphics[width=0.48\linewidth]{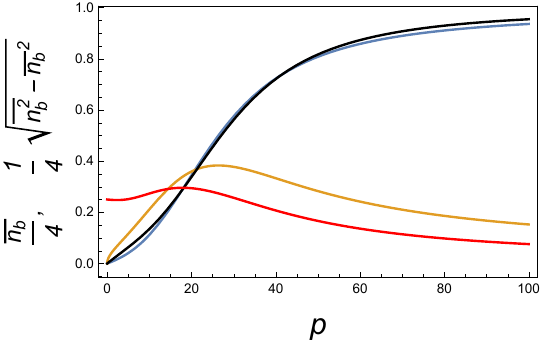}
    \caption{Predictions of Pauling's and the MWC model (grand canonical ensemble) versus the partial pressure $p$. Blue and orange curves are (resp.) the mean saturation $\frac{1}{4}\overline{n_b}$ and the standard deviation $\frac{1}{4}\sqrt{\overline{n_b^2}-\overline{n_b}^2}$ calculated from (\ref{lulu}) under parameters (\ref{bard1}) reported by Pauling \cite{pauling_pnas}. Black and red curves are (resp.) the mean saturation $\frac{1}{4}\overline{n_b}$ and the standard deviation $\frac{1}{4}\sqrt{\overline{n_b^2}-\overline{n_b}^2}$ calculated for the MWC model [cf.~(\ref{bru0})] under parameters (\ref{bard3}) reported by Eaton \cite{eaton_models}. For the MWC model, we do not show $\pi({\rm R})$ (probability of the R-state), since for the considered range of parameters it nearly coincides with $\frac{1}{4}\overline{n_b}$.  }
    \label{fig-2}
\end{figure}

The mean number of bound particles $\overline{n_b}$ found from (\ref{hmean}, \ref{cosmo}) holds [see Table~\ref{tab2}]:
\BEA
\overline{n_b}\simeq n\gg \sqrt{\overline{n_b^2}-\overline{n_b}^2}, 
\EEA
i.e., \O molecules are strongly bound to the Hb. In contrast, for the grand canonical ensemble of Pauling's model, we get $\overline{n_b}\sim \sqrt{\overline{n_b^2}-\overline{n_b}^2}$ indicating on the relatively weak binding; see Fig.~\ref{fig-2}. Another difference between canonical and grand canonical situations is that for the former, $\sqrt{\overline{n_b^2}-\overline{n_b}^2}$ grows monotonically with $n$; see Table~\ref{tab2} and compare it with Fig.~\ref{fig-2}. 

Why is cooperativity needed at all? For $\alpha=1$ (no cooperativity) and $K\simeq 200$ we can satisfy (\ref{test4}), but (\ref{test3}--\ref{test1}) cannot be satisfied, because the corresponding values of $|g[3]|$, $|g[2]|$ and $|g[1]|$ are too big; see Fig.~\ref{fig-3}. Hence, without cooperativity the Hb can be loaded in lungs, but will serve only resting tissues; see Table~\ref{tab1}.

What can happen for such large values of $\alpha$?
Fig.~\ref{fig-3} shows $\{g[n]\}_{n=1}^4$ for $K=3.7$ and $\alpha=7.83$. The latter value is chosen such that inequality (\ref{test4}) is satisfied. It is seen that the basic condition (\ref{c1}) does not hold for $n=3,2,1$, i.e., such a model Hb cannot be loaded in lungs. We can take somewhat smaller value of $\alpha$ such that (\ref{c1}) is satisfied: see Fig.~\ref{fig-3} for $K=13$ and $\alpha=4.83$. 
But now we get a cascade process of \O release, which is described via
\BEA
\label{cascade}
g[4]<g[n],\quad n=3,2,1.
\EEA
This means that once the fourth bound \O molecule is released -- e.g., for $g[4]>5.1+\ln[10/8]$, as in (\ref{test4}) -- the other three molecules will be released automatically. This is not an acceptable scenario, because it will lead to \O poisoning. 

\begin{table}[t]
\caption{For Pauling's model under (\ref{vard1}), this table shows that $\overline{n_b}\simeq n\gg \sqrt{\overline{n_b^2}-\overline{n_b}^2}$, i.e., \O molecules are strongly bound to the Hb macromolecule. For the grand canonical ensemble of Pauling's model, we get 
$\overline{n_b}\sim \sqrt{\overline{n_b^2}-\overline{n_b}^2}$ indicating on the relatively weak binding; see
Fig.~\ref{fig-2}. Note that the data in this table shows that $\sqrt{\overline{n_b^2}-\overline{n_b}^2}$ grows with $n$, in contrast to grand canonical behavior of $\sqrt{\overline{n_b^2}-\overline{n_b}^2}$. Also $\overline{n_b}\Big/\sqrt{\overline{n_b^2}-\overline{n_b}^2}$ grows with $n$ changing from 16.9 for $n=1$ to 31.9 for $n=4$.
}
\begin{center}
\begin{tabular}{|c|| c | c | c | c |} 
 \hline
 $n$                       & 4   &   3 &   2 & 1 \\ 
 \hline \hline
 $\overline{n_b}$, & 3.984 & 2.998 & 1.996 & 0.997  \\ 
 \hline
 $\sqrt{\overline{n_b^2}-\overline{n_b}^2}$ & 0.125 & 0.097 & 0.077 & 0.0590 \\ 
 \hline
\end{tabular}
\end{center}
\label{tab2}
\end{table}

\begin{figure}[t]
    \centering
    \includegraphics[width=0.48\linewidth]{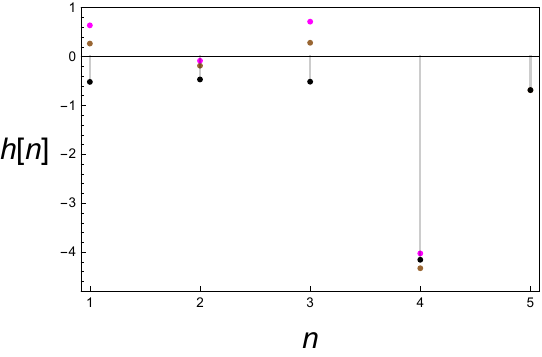}
    \caption{$h[n]$ versus $n$ for Pauling's model; see (\ref{baran}, \ref{pau2}). Black, magenta, and brown points are the values of $h[n]$ for (resp.) ($K=71.33$, $\alpha=1.88$), ($K=3.7$, $\alpha=7.83$), and ($K=13$, $\alpha=4.83$); cf.~(\ref{vard1}, \ref{bard2}). Now $h[n]>0$ means that the Hb pair with $(n;n)$ \O molecules will spontaneously move (due to cooperativity) to $(n+1;n-1)$; cf.~(\ref{baran0}).  }
    \label{fig-4}
\end{figure}

Fig.~\ref{fig-4} shows for Pauling's model the free energy difference $h[n]$ defined in (\ref{baran}). It is seen that for all studied values of parameters, the $(n=4;n=4)$ state of the Hb pair is the most stable one against the spontaneous gradient formation: $h[4]$ assumes the largest value, which is almost parameter independent; see Fig.~\ref{fig-4}. However, for strongly cooperative situations with $\alpha=4.83$ and $\alpha=7.83$, the states $(n=3;n=3)$ and $(n=1;n=1)$ are instable: $(n=3;n=3)$ spontaneously moves to $(n=4;n=2)$, while $(n=1;n=1)$ moves to $(n=1;n=0)$. Though we do not involve detailed dynamical calculations in this paper, we note that such spontaneous gradient formations will impede the \O diffusion across the RBC. 

\comment{We assume that $n$ is large in (\ref{h1}, \ref{z}, \ref{hmean}): $k!\, C^k_n V^{-k}\simeq (n/V)^k$. Hence only the overall concentration
enters, which can also be represented via the ideal-gas law and partial \O pressure $p\beta =n/v$
and use (\ref{h3}) and the partial pressure $pV=nT$:
\BEA
\label{h7}
p(k)=\frac{1}{Z}C^k_4 p^{k}\prod_{i=1}^k \beta v(i) e^{-\beta \varepsilon(i)}.
\EEA
}

\subsection{Monod-Wyman-Changeux (MWC) model within the canonical ensemble}
\label{monod}

We return to (\ref{zzz}) and recall that this model describes cooperativity via 
direct coupling between binding sites. Within (\ref{zzz}), no cooperativity means 
that $f(i)=f$ does not depend on $i$, and then $\prod_{i=0}^k e^{-\beta f(i)}=e^{-\beta k f}$.
In addition to this assumption of no direct cooperativity, MWC model introduces two different states (R and T), which have different affinity to oxygen \cite{eaton_models,mwc,changeux_review_50}; see (\ref{yak}) below. Now $f$ depends on the state: $f(s)$, $s={\rm R, T}$. The mechanism of cooperativity proposed by MWC allows thermal transitions between T and R, and chooses $f(s)$ such that these transitions are influenced by oxygen binding. In this way, MWC model explains cooperativity via allostery \cite{bellelli}.
Like Pauling's model, MWC is based on statistical thermodynamics. According to Perutz, the MWC model is based on conformational features (quaternary and partially tertiary) of Hb: R (T) refers to oxyhemoglobin (deoxyhemoglobin); see \cite{perutz} for a review.  

The canonical partition function of the MWC model is found by analogy to (\ref{zzz}): 
\BEA
{\cal Z}=\frac{V^n}{n!} 
\sum_{s={\rm R, T}}
e^{-\beta E(s)}\sum_{k=0}^4 k! \binom{n}{k}\binom{4}{k}
\,e^{-\beta k f(s)}=\frac{V^n}{n!} Z,
\label{orosh}
\EEA
where the energy $E(s)$ controls relative weights of R and T states. Working out (\ref{orosh}) we get
\BEA
\label{zm}
Z[n]&=&(L+1)+4n (LK_{\rm T}+K_{\rm R})+6n(n-1)(LK_{\rm T}^2+K_{\rm R}^2)
\nonumber\\
&&+4n(n-1)(n-2)(LK_{\rm T}^3+K_{\rm R}^3)+n(n-1)(n-2)(n-3)(LK_{\rm T}^4+K_{\rm R}^4),\\
L&=&e^{-\beta E(T)}, \qquad K_{\rm T}=e^{-\beta f(T)}, \qquad K_{\rm R}=e^{-\beta f(R)}, \qquad E(R)=0,
\label{zm2}
\EEA
where choosing $E(R)=0$ does not contain any loss of generality. The probabilities $p(k)$ of $k$ bound \O molecules follow from (\ref{zm}) and refer to separate terms in $Z$:
\BEA
p(0)=\frac{L+1}{Z},\quad p(1)=\frac{4n (LK_{\rm T}+K_{\rm R})}{Z},\quad 
p(2)=\frac{6n(n-1)(LK_{\rm T}^2+K_{\rm R}^2) }{Z},...
\comment{
&&p(3)=\frac{4n(n-1)(n-2)(LK_{\rm T}^3+K_{\rm R}^3) }{Z},\quad p(4)=\frac{n(n-1)(n-2)(n-3)(LK_{\rm T}^4+K_{\rm R}^4) }{Z}.
}
\label{baghdad}
\EEA
Likewise, the probability of the $R$-state reads
($\pi({\rm R})+\pi({\rm T})=1$)
\BEA
\pi({\rm R})&=&  \frac{e^{-\beta E({\rm R})}}{Z} 
\sum_{k=0}^4 k! C^k_nC^k_4 \,e^{-\beta k f({\rm R})}\nonumber\\
\label{bru}
&=&\frac{1}{Z}[1+4n K_{\rm R}+6n(n-1)K_{\rm R}^2+4n(n-1)(n-2)K_{\rm R}^3
+n(n-1)(n-2)(n-3)K_{\rm R}^4].
\comment{\frac{\pi(R)}{\pi(T)}=\frac{1}{L}\,\,\,
\frac{1+4n K_{\rm R}+6n(n-1)K_{\rm R}^2+4n(n-1)(n-2)K_{\rm R}^3
+n(n-1)(n-2)(n-3)K_{\rm R}^4}
{1+4n K_{\rm T}+6n(n-1)K_{\rm T}^2+4n(n-1)(n-2)K_{\rm T}^3
+n(n-1)(n-2)(n-3)K_{\rm T}^4}.
\label{karp}
}
\EEA
For three limiting cases: $K_{\rm T}=K_{\rm R}$, $L=0$, and $L=\infty$ we get from (\ref{orosh}, \ref{bru}) the (simplest) non-cooperative version of the Hb-\O equilibrium with a single parameter $K$. More generally, the model has three independent parameters: $K_{\rm T}$, $K_{\rm R}$, and $L$, and the deviation from the non-cooperative equilibrium is not characterized by a single parameter. 

We now calculate $g[n]$ via (\ref{zm}, \ref{mree}). The best fit of parameters $K_{\rm T}>0$, $K_{\rm R}>0$, and $L>0$ to the data in Table~\ref{tab1} [see before (\ref{vard1}) for details] produced:
\BEA
K_{\rm R}=331.7, \quad K_{\rm T}=70.7, \quad L=39.2.
\label{yak}
\EEA
For the grand-canonical model the fit to the data of Table~\ref{tab1} led to a sizable difference: $K_{\rm R}/K_{\rm T}=95.4$ and $L=8.26\times 10^5$; see (\ref{bard3}, \ref{bard4}) in Appendix \ref{apo}. The latter value is comparable with the value $L=5\times 10^5$ reported by Eaton \cite{eaton_models}. As compared to (\ref{yak}), these grand-canonical values indicate a stronger deviation from the non-cooperative situation.

In contrast to Pauling's model, where there is a single parameter $\alpha$ characterizing direct interaction between binding sites, the cooperativity in the MWC model is an emergent feature of the equilibrium shift due to \O binding. Eqs.~(\ref{bru}, \ref{yak}) show that for $n=1$ we have $\frac{\pi(R)}{\pi(T)}=\frac{1}{L}\,\frac{1+4K_{\rm R}}{1+4K_{\rm T}}\simeq 0.11$, while for $n=4$ we have $\frac{\pi(R)}{\pi(T)}=\frac{K_{\rm R}^4  }{L K_{\rm T}^4}\gg 1$ due to 
\BEA
(K_{\rm R}/K_{\rm T})^4\gg L\gg 1,
\label{dedekind}
\EEA
which is satisfied for (\ref{yak}). Table~\ref{tab3} shows results found from the MWC model under (\ref{yak}). The behavior of $\overline{n_b}$ and $\sqrt{\overline{n_b^2}-\overline{n_b}^2}$ 
is very similar to what Table~\ref{tab2} predicts for Pauling's model. Again, we witness a clear difference with the grand canonical situation. As expected, $\pi(R)$ increases from $\sim 0.1$ with one \O binding to $\sim 1$ when all four \O molecules are bound. Given the deterministic behavior of the number of bound \O molecules [see Table~\ref{tab3}], and indeterministic behavior of $\pi(R)$, it is legitimate to say that the \O binding indeed drives the state change. This contrasts the grand canonical situation, where the number of bound particles and the state are two correlated random variables.  

The behavior of $g[n]$ and $h[n]$ for the MWC model under the best fit (\ref{yak}) is again similar to Pauling's model under parameters (\ref{vard1}) (which is the best fit for Pauling's model); see Figs.~\ref{fig-3} and \ref{fig-4}.
Hence, we do not present them in a separate figure. 

\subsubsection{Instability for large cooperativity}

The instability effects noted for Pauling's model in the stronger cooperativity regime are also present here; cf.~(\ref{cascade}). We illustrate those instabilities as follows. 
We apply the same best fit that led to (\ref{yak}), but now we fix $L$ to a value $L=8\times 10^5$, which is found from the grand-canonical ensemble; see (\ref{bard4}) in Appendix \ref{apo}. Now we find 
\BEA
K_{\rm R}=3577.9, \quad K_{\rm T}=104.7, \quad L=8\times 10^5,
\label{dode}
\EEA
which satisfies (\ref{dedekind}). Note that the ratio $K_{\rm R}/K_{\rm T}=34.2$ deduced from (\ref{dode}) (recall that this ratio characterizes the deviation from the non-cooperative situation) is still lower than the ratio
$K_{\rm R}/K_{\rm T}=95.4$ obtained as the best to the grand canonical ensemble; see (\ref{bard4}) in Appendix \ref{apo}.

Parameters (\ref{dode}) lead to stability conditions $h[n]<0$ for $n=1,..,4$; see (\ref{baran0}). 
However, (\ref{dode}) does not satisfy (\ref{test3}) due to $g[3]\approx g[4]$, but still $g[3]<g[4]$. Hence, under (\ref{dode}), the third bound \O molecule is released almost simultaneously with the fourth one. Conditions (\ref{test4}, \ref{test2}, \ref{test1}) are satisfied with (\ref{dode}). 

For parameters $K_{\rm R}=3577.9$, $K_{\rm T}=104.7$, and $L=6\times 10^5$, we get $g[3]> g[4]$, which is similar to the instability in (\ref{cascade}). Also, for this situation, there is another sign of instability: $h[3]>0$, while $h[1]$, $h[2]$ and $h[4]$ are negative.

Thus, also for MWC model, parameters inferred from the grand canonical situation lead to an instability in the canonic setting. 

\begin{table}[t]
\caption{For MWC model under (\ref{yak}), this table shows 
$\overline{n_b}$, $\sqrt{\overline{n_b^2}-\overline{n_b}^2}$, and $\pi(R)$ versus $n$. We see that
$\overline{n_b}\simeq n\gg \sqrt{\overline{n_b^2}-\overline{n_b}^2}$, i.e., \O molecules are strongly bound to the Hb macromolecule. For the grand canonical ensemble of MWC model, we get 
$\overline{n_b}\sim \sqrt{\overline{n_b^2}-\overline{n_b}^2}$ indicating on the relatively weak binding; see
Fig.~\ref{fig-2}. The first two lines of this table are close to the first two lines of Table~\ref{tab2}. It is seen that the R-state dominates (i.e., $\pi({\rm R})>0.5$) after binding the third oxygen molecule.
Note that for the grand canonical implementation of the MWC model, $\pi({\rm R})$ and 
$\frac{1}{4}\overline{n_b}$ are very close to each other; cf.~Fig.~\ref{fig-2}. In contrast, for the canonical situation presented in this table, $\frac{1}{4}\overline{n_b}$ is visibly larger than $\pi({\rm R})$. 
}
\begin{center}
\begin{tabular}{|c|| c | c | c | c |} 
 \hline
 $n$                       & 4   &   3 &   2 & 1 \\ 
 \hline \hline
 $\overline{n_b}$, & 3.985 & 2.991 & 1.993 & 0.997  \\ 
 \hline
 $\sqrt{\overline{n_b^2}-\overline{n_b}^2}$ & 0.123 & 0.095 & 0.082 & 0.057 \\ 
 \hline
 $\pi({\rm R}) $ & 0.922 & 0.722 & 0.358 & 0.107 \\ 
 \hline
\end{tabular}
\end{center}
\label{tab3}
\end{table}

\begin{table}[t]
\caption{For MWC model under  $K_{\rm R}=1494.9$ and  $K_{\rm T}=316.1$ [cf.~(\ref{yak})], this table shows $g[n]$ under various values of  $L$. 
}
\begin{center}
\begin{tabular}{|c|| c | c | c | c |} 
 \hline
 $n$                       & 4   &   3 &   2 & 1 \\ 
 \hline \hline
 $L=42.3$             & -5.67  & -6.18  & -6.48  & -7.22  \\ 
 \hline\hline
 $L=52.3$             & -5.62  & -6.10  & -6.43  & -7.21  \\ 
 \hline
 $L=72.3$             & -5.54  & -5.98  & -6.37  & -7.19  \\ 
 \hline
 $L=102.3$             & -5.43  & -5.86  & -6.31  & -7.18  \\ 
 \hline
\end{tabular}
\end{center}
\label{tab5}
\end{table}

\begin{table}[t]
\caption{
For MWC model under  $K_{\rm R}=1494.9$ and  $L=42.3$ [cf.~(\ref{yak})], this table shows $g[n]$ under various values of  $K_{\rm T}$. 
}
\begin{center}
\begin{tabular}{|c|| c | c | c | c |} 
 \hline
 $n$                       & 4   &   3 &   2 & 1 \\ 
 \hline \hline
 $K_{\rm T}=316.1$ & -5.67  & -6.18  & -6.48  & -7.22   \\ 
 \hline\hline
 $K_{\rm T}=500.$ & -5.40  & -6.11  & -6.74  & -7.65 \\ 
 \hline
 $K_{\rm T}=750.$ & -5.38  & -6.30  & -7.07  & -8.03  \\ 
 \hline
\end{tabular}
\end{center}
\label{tab4}
\end{table}

\subsubsection{The influence of ligands (within MWC model)}

Hb ligands are of two types: heme ligands (e.g., \O, CO, or NO) bind to the heme pocket. Non-heme ligands (or heterotropic ligands; e.g. protons ${\rm H^+}$, 2,3-BPG, and ${\rm CO_2}$) bind outside of the heme pocket. Normally, binding of heme ligands decreases the affinity for non-heme ligands and vice versa \cite{bellelli}. A related empiric rule is that heme ligands bind with higher affinity to the R state, while non-heme ligands usually bind with higher affinity to T \cite{bellelli}.

Heterotropic ligands 2,3-BPG and proton (${\rm H^+}$) play important roles in maintaining and regulating transport functions of Hb \cite{perutz,ahmed,bellelli}. In particular, they facilitate \O release in capillaries \cite{eaton_models}. The concentration $n_{\rm 2,3-BPG}$ of 2,3-BPG in RBC changes under various physiological and clinical conditions, since this concentration regulates \O binding \cite{perutz, ahmed, duhm}. Increasing $n_{\rm 2,3-BPG}$, which is frequent under anemia and hypoxia, facilitates the \O unbinding from Hb. Decreasing $n_{\rm 2,3-BPG}$, which happens after blood conservation, strengthens the \O binding to Hb \cite{duhm}. The proton binding to Hb is central for Bohr's effect \cite{bellelli}.

The influence of 2,3-BPG and ${\rm H^+}$ on Hb-\O equilibrium was already studied within the grand-canonical MWC model \cite{eaton_models}. They bind to the T-state and thereby influence mostly parameters $K_{\rm T}$ and $L$ \cite{perutz}. Ref.~\cite{mwc} initially suggested that only $L$ is affected, but this assumption turned out to be too restrictive \cite{eaton_models}. Within MWC model, these ligands bias the allosteric equilibrium of the model towards the T state and hence facilitate \O release \cite{eaton_models}. 

We now show that the canonical MWC model is consistent with the expected behavior of the HB-\O equilibrium under the influence of heterotropic ligands 2,3-BPG and ${\rm H^+}$. Note that ligands were already taken into account in the data of Table~\ref{tab1} by stating standard physiological conditions. 
Therefore, we mean conditions that go beyond physiological ones.

We want to understand how $g[n]$ [see (\ref{zm}, \ref{mree})] behaves under changing $K_{\rm T}$ and $L$ (while $K_{\rm R}$ is fixed). Table~\ref{tab5} shows that upon increasing $L$, $g[n]$ monotonically increases, i.e. the \O release is facilitated. This is the same conclusion, as reached under increase of $L$ in the grand canonical ensemble; see Appendix \ref{apo-mwc}. The behavior under increasing $K_{\rm T}$ is more convoluted, as Table~\ref{tab4} shows: $g[4]$ increases (\O release facilitation), $g[3]$ changes non-monotonically, whereas $g[2]$ and $g[1]$ decrease thereby stabilizing the \O binding more. This behavior does not contradict results obtained from the grand canonical ensemble; see Appendix \ref{apo-mwc}.


\section{Summary}

Our main message is that the distribution of oxygen (\O) and hemoglobin (Hb) in RBC (red blood cells) require changing the approach of statistical physics to the  Hb-\O equilibrium.
The usual approach is based on the grand canonical ensemble, which models a single Hb immersed into a reservoir of many \O molecules, where the reservoir is described by the concentration (or partial pressure) of \O \cite{dill}. As we argued in section \ref{intro}, this is not the situation realized in RBC, where each Hb most probably sees $n<5$ \O molecules. A better model is to assume that a single Hb with a few \O molecules surrounding it form a canonical ensemble; see Table~\ref{tab11}.

In this model, {thermodynamic} features of oxygen transport are determined via free energy differences discussed in section \ref{frea}. The free energy difference is conceptually distinct from the mean number of bound \O molecules, which (as a function of the \O partial pressure) determines the Hb-\O equilibrium within the grand canonical ensemble; see Table~\ref{tab-ra} for a detailed comparison between the two ensembles. 
The above distinction between the ensembles leads to contrasting predictions. We examined them in two basic models that describe the Hb-\O equilibrium and provide different explanations for cooperativity: Pauling's model and MWC model. It turns out that the cooperativity values predicted by the grand canonical ensemble are not suitable for the canonical situation, because they will lead to various forms of instability discussed in Figs.~\ref{fig-3} and \ref{fig-4}, as well as around (\ref{dode}). 
Also, Hb behaves differently in each ensemble. Within the grand canonical ensemble, the number of bound \O molecules is a fluctuating quantity, whose mean follows the saturation data of Fig.~\ref{tab1}. Within the canonical approach, every available \O molecule is bound, and once Hb cannot bind \O anymore, it also does not attract them; see the discussion after (\ref{kau}).  

Anticipating future developments, we note that our results on the thermodynamics of \O diffusion indicate that \O can be transported within RBC via hopping diffusion along with the usual diffusion; for recent results in oxygen diffusion inside RBC see \cite{richardson_pnas}. To design nonequilibrium models of this type, one should fill the RBC volume with cubes of volume $V$, each one containing a single hemoglobin; cf.~Fig.~\ref{fig-1}. Hopping diffusion means that \O molecules move from one cube to another also taking into account effective thermodynamic forces that arise due to free energy differences; see section \ref{tramspo}. The hypothesis should be corroborated in future dynamical studies, also taking into account both rotational and translational diffusion of Hb macromolecules; for reviews on hopping diffusion see \cite{richards,derrida,collins}. 

\acknowledgements

We thank Arik Avagyan for discussions. The work was supported by the HESC of Armenia, grant No. 21AG-1C038. 

\section*{Conflict of Interest Statement}

The authors have no conflicts to disclose.

\section*{Author Contributions (CRediT Taxonomy)}

AEA: Conceptualization, Formal analysis, Writing – Original Draft;

SGS: Conceptualization, Writing – Review \& Editing;

AH: Data Curation, Methodology, Software, Writing – Review \& Editing;

YShM: Validation, Investigation, Writing – Review \& Editing;

VS: Conceptualization, Formal analysis, Writing – Review \& Editing;

\section*{Data Availability Statement}

The data are within the article. 

\comment{Significance statement

This study provides a new perspective on the hemoglobin-oxygen equilibrium, arguably one of the most important problems in modern biological physics and its applications. It should be of interest to all physicists interested in biology, as well as to chemical physicists and chemists who have a working knowledge of statistical physics.

}

\appendix

\section{ Grand canonical ensemble}
\label{apo}

Here we review the grand canonical formulation of Pauling's and MWC model, and determine their parameters by fitting to data of Table~\ref{tab1}. 

\subsection{Pauling's model }

We apply the procedure described in the beginning of section \ref{grando} to (\ref{pau2}), and introduce a new parameter $K'$ ($K$ is defined before (\ref{pau1})):
\BEA
\label{otk}
Kn=e^{-\beta \varepsilon(1)}v(1)\frac{n}{V}=K'p. 
\EEA
Eq.~(\ref{otk}) illustrates the essence of the grand canonical ensemble: for $n\gg 1$, only $n/V\propto p$ enters observable quantities. We get from (\ref{pau2}) for the mean number $\overline{n_b}$ of bound \O molecules, and the variance $\overline{n_b^2}-\overline{n_b}^2$ \cite{pauling_pnas}:
\BEA
&& Z=1+4pK'+ (4\alpha+2)p^2 K'^2+4p^3\alpha^2 K'^3+ p^4\alpha^4 K'^4,\\
&& \overline{n_b}=p\frac{\partial}{\partial p}\ln Z, \qquad 
\overline{n_b^2}-\overline{n_b}^2=p\frac{\partial}{\partial p}\overline{n_b}.
\label{lulu}
\EEA
The mean number $\overline{n_b}$ in (\ref{lulu}) as a function of $p$ (with parameters $K'$ and $\alpha$)
is the main outcome of the grand canonical approach, because $\frac{1}{4}\overline{n_b}$ versus $p$ can be compared with saturation curves \cite{pauling_pnas}. This determines $K'$ and $\alpha$ from curve-fitting \cite{pauling_pnas}. 
Pauling employed the data by Ferry and Green \cite{ferry} and found from the best fit
\cite{pauling_pnas}:
\BEA
\label{bard1}
K'=0.0033,\quad \alpha=12.
\EEA
We used the coarser data from Table~\ref{tab1}, which represents $\frac{1}{4}\overline{n_b}$ 
as a function of $p$, and fitted it to $\frac{1}{4}\overline{n_b}$ in (\ref{lulu}) to find parameters $K'$
and $\alpha$. The best fit values are:
\BEA
\label{bard2}
K'=0.006,\quad \alpha=6.32.
\EEA
The values of $K'$ in (\ref{bard1}) and (\ref{bard2}) are numerically comparable to each other. 
Though the values of $\alpha$ in (\ref{bard1}) and (\ref{bard2}) are numerically different, they produce qualitatively the same results if employed for the canonical model; see Fig.~\ref{fig-3}. 
Recall in this context that the parameter $\alpha$ has the same form and the same meaning for both 
canonical and grand canonical ensembles.

\subsection{MWC model}
\label{apo-mwc}

For $n\gg 1$ we get from (\ref{zm})
\BEA
\label{bru0}
Z=L(1+K_{\rm T}n)^4+(1+K_{\rm R}n)^4,
\EEA
which is the standard partition function of the MWC model \cite{eaton_models}. 
We carry out the same procedure as before (\ref{otk}):
\BEA
K_{\rm T}n=K_{\rm T}'p, \qquad K_{\rm R}n=K_{\rm R}'p,
\label{mango}
\EEA
and employ (\ref{lulu}) for MWC model. We get for the probabilities of T and R states:
\BEA
\pi({\rm R})=\frac{1}{Z}[1+K_{\rm R}'p]^4 , \qquad
\frac{\pi({\rm R})}{\pi({\rm T})}=\frac{1}{L}\,\,\,
\frac{[1+K_{\rm R}'p]^4}
{[1+ K_{\rm T}'p ]^4}.
\label{bru2}
\EEA

\begin{figure}[t]
    \centering
    \includegraphics[width=0.48\linewidth]{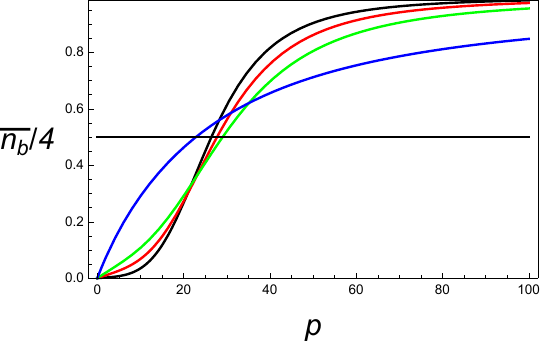}
    \caption{For MWC model we plot $\frac{1}{4}\overline{n_b}$ versus the partial pressure $p$ (measured in torr) for $K'_{\rm R}=1$ and $L=5\times 10^5$; see (\ref{bard3}). Black curve $K'_{\rm T}=0.001$, red $K'_{\rm T}=0.005$, green $K'_{\rm T}=0.01$, blue $K'_{\rm T}=0.04$. It is seen that $p_{50}$ (the partial pressure, where $\frac{1}{4}\overline{n_b}(p_{50})=0.5$) changes non-monotonically upon increasing $K'_{\rm T}$.   
    }
    \label{fig-5}
\end{figure}

\comment{
\BEA
\label{blue1}
&&p(0)=\frac{L+1}{Z},\quad p(1)=\frac{4n (LK_{\rm T}+K_{\rm R})}{Z},\quad 
p(2)=\frac{6n^2(LK_{\rm T}^2+K_{\rm R}^2) }{Z},\\ 
&&p(3)=\frac{4n^3 (LK_{\rm T}^3+K_{\rm R}^3) }{Z},\quad p(4)=\frac{n^4 (LK_{\rm T}^4+K_{\rm R}^4) }{Z},\\
&&\pi(R)=\frac{1}{Z}[1+n K_{\rm R}]^4 , \qquad
\frac{\pi(R)}{\pi(T)}=\frac{1}{L}\,\,\,
\frac{[1+n K_{\rm R}]^4}
{[1+n K_{\rm T}]^4}.
\label{bru2}
\EEA
}
Eqs.~(\ref{bru0}--\ref{bru2}) are studied within conditions (\ref{dedekind}), where $K_{\rm R}/K_{\rm T}=K_{\rm R}'/K_{\rm T}'$, as seen from (\ref{mango}). Conditions (\ref{dedekind}) ensure the following scenario. For low \O concentration 
$pK_{\rm T}'\ll pK_{\rm R}'\lesssim 1$ we get from (\ref{bru2}, \ref{dedekind}):
$\bar{n}\to 0$ and $\pi(T)\gg \pi(R)$.
This means that for low \O concentration, no \O is bound and the T-state is dominant. 
This is consistent with the fact $E(T)$ is negative, while $E(R)$ is zero; see (\ref{zm}). For higher $p$,
$pK_{\rm R}'\gg pK_{\rm T}'> 1$ we get from (\ref{bru2}, \ref{dedekind}):
$\bar{n}\sim 4$ and $\pi(T)\ll \pi(R)$.
Now the $R$-state is dominant and the Hb is fully loaded. Between high and 
low $p$ regimes, the mean number of bound \O molecules changes via a S-shaped curve.

Note that the values of $K_{\rm R}'/K_{\rm T}'$ and $L$ for the grand canonical situation can be compared to (resp.)
$K_{\rm R}/K_{\rm T}$ and $L$ for the canonical ensemble. 
Eaton reports the following values found from the best-fit to several datasets \cite{eaton_models}:
\BEA
\label{bard3}
K_{\rm R}'=1.,\quad K_{\rm T}'=0.001,\quad L=5\times 10^5.
\EEA
We used the data from Table~\ref{tab1} and found from the best fit
a larger value for $L$, but a smaller value for the ratio $K_{\rm R}/K_{\rm T}$: 
\BEA
\label{bard4}
K_{\rm R}'=1.24,\quad K_{\rm T}'=0.013,\quad L=8.26\times 10^5.
\EEA
Both (\ref{bard3}) and (\ref{bard4}) satisfy (\ref{dedekind}). 

Let us now see how the basic diagnostic quantity of the grand canonical ensemble MWC model, i.e. $\overline{n_b}$ reacts on changes of $K'_{\rm T}$ and $L$. We restrict ourselves by these two parameters, because heterotropic ligands (2,3-BPG and protons) specifically bound in the T-state, and are standardly assumed to influence only $K'_{\rm T}$ and $L$ \cite{perutz}. Note that from (\ref{lulu}, \ref{bru0})
\BEA
\frac{\partial}{\partial L}\overline{n_b}=
\frac{4 p}{Z^2} (K'_\text{T}-K'_\text{R}) (K'_\text{R} p+1)^3 (K'_\text{T} p+1)^3,
\EEA
which means that in the considered regime $K'_{\rm R}>K'_{\rm T}$, increasing $L$ will decrease 
$\overline{n_b}$. 

The response of $\overline{n_b}$ to changes of $K'_{\rm T}$ is more convoluted: increasing $K'_{\rm T}$ increases $\overline{n_b}$ for smaller values of $p$, but decreases it for larger values of $p$; see Fig.~\ref{fig-5}. Put differenty, increasing $K'_{\rm T}$ facilitates \O release, but only for larger values of $p$. In this context, one frequently involves 
$p_{50}$, which the partial \O pressure, where $\frac{1}{4}\overline{n_b}(p_{50})=0.5$. 
Fig.~\ref{fig-5} shows that $p_{50}(K'_{\rm T})$ changes non-monotonically upon increasing $K'_{\rm T}$.

\section{Free energy differences when transferring more than 2 particles}
\label{2molecules}

\subsection{Transfer of $s$ molecules}

We can generalize (\ref{free}, \ref{mree}) for the case when $s$ particles are transferred:
\BEA
&&\Delta {F}_{\rm tot}[s]\equiv { F}_{\rm tot}[n,\hat N]-{ F}_{\rm tot}[n-s,\hat N+s]
=Ts\Big(g[n,s]-\ln\frac{V\hat N}{ \hat V }\Big),
\label{frees}
\\
&&g[n,s]\equiv \frac{1}{s}\ln\frac{Z[n-s]n!}{Z[n](n-s)!}.
\label{mrees}
\EEA
where in (\ref{frees}) we assumed that $\hat N\gg 1$. Now $\Delta {F}_{\rm tot}[s]<0$ in (\ref{free}) means that $s$ \O molecules are absorbed by the Hb, while $\Delta {F}_{\rm tot}[s]>0$ means that $s$ \O molecules are released by the Hb. 

We saw in (\ref{kau}) that one \O transfer from the blood to the fully loaded Hb molecule is not possible; i.e. $g[5,1]>0$ in notations of (\ref{mrees}). It turns out that 2-oxygen transfer is also impossible. For Pauling's model, we get employing (\ref{vard1}): 
\BEA
\label{bbamba}
g[5,2]=-2.82,\quad g[6,2]=0.348,\quad g[7,2]=0.896. 
\EEA
Even though now $g[5,2]<0$, it is not sufficiently negative to satisfy (\ref{c1}). 
Changing parameters in (\ref{vard1}) does not affect (\ref{bbamba}) at least within the presented order of magnitude. Eq.~(\ref{bbamba}) holds also for MWC model, where it is also nearly parameter independent.

\subsection{Transfer of one molecule in a larger gradient}

Eq.~(\ref{baran}) can be generalized to one \O molecule transfer, when the \O gradient between two Hb molecules amounts to 3 molecules:
\BEA
&&F[n]+F[n+1] - (F[n+2]+F[n-1])=Th_2[n],\\
&&h_2[n]\equiv \ln\frac{nZ[n+2]Z[n-1]}{Z[n+1] Z[n](n+2)},
\EEA
where $h_2[n]<0$ means tendency towards gradient decreasing. For a strong cooperativity, $h_2[n]$ shows the same type of pathologies as $h[n]$. For example, in Pauling's model with $K=15$ and $\alpha=7.83$, we get 
$h_2[2]=0.594$ and $h_2[1]=0.614$, while $h_2[n\geq 3]<0$. For $K=56.$ and $\alpha=4.83$, the same picture holds but with smaller absolute values of positive terms: $h_2[2]=0.0.086$ and $h_2[1]=0.09$, and $h_2[n\geq 3]<0$. For (\ref{vard1}) we get $h_2[n\geq 1]<0$ indicating on thermodynamic stability.

\section{Full (coordinate plus momentum) partition function and free energy}
\label{apo_full}

In this Appendix, we determine the momentum part of the partition function to be added to the coordinate part (\ref{zzz}). For $n$ particles the momentum (Maxwell's) distribution leads to the following total partition function $Z_{\rm total}$ and the corresponding free energy $F_{\rm total}$ \cite{landau5}:
\BEA
\label{gibbi}
F_{\rm total}=-T\ln {\cal Z}_{\rm total}=-T\ln Z_{\rm momenta}-T\ln {\cal Z}=
-\frac{3nT}{2}\ln\Big[ \frac{mT}{2\pi\hbar} \Big]-T\ln{\cal Z}.
\EEA
For pressure $P$ and the mean energy $U$ we have from (\ref{gibbi}, \ref{zzz}):
\BEA
\label{chev}
&& P=-\frac{\partial F_{\rm total} }{\partial V}=\frac{T}{V}(n- \overline{n_b}  ), \\
&& U=F_{\rm total}+TS=F_{\rm total}-T\frac{\partial F_{\rm total} }{\partial T}  =\frac{3nT}{2}
-\frac{\partial }{\partial\beta}\ln Z.
\label{tukh}
\EEA
Both these formulas have transparent meaning: (\ref{chev}) means that the pressure is created by unbound particles (as it should), while (\ref{tukh}) means that the mean energy is determined from $\ln Z$ plus the mean kinetic energy. 

\section{Possible generalizations of the grand canonical ensemble}
\label{apo_clarifo}

The above derivation of the grand canonical ensemble (presented in section \ref{grando}) assumes one Hb macromolecule and many $n\gg 1$ \O molecules. 
Given that the ensemble does not apply to RBC, one wonders whether the grand canonical ensemble can still apply to cases in which many Hb molecules interact with many \O molecules. To answer this question, let us now assume that we have $\Nhb>n$ for $\Nhb$ molecules of Hb and $n$ molecules of \O. This situation is realized in RBC; see Table \ref{tab11}. We now demonstrate that the grand canonical ensemble (discussed in section \ref{grando} and in Appendix \ref{apo}) does not emerge even in the limit $\Nhb>n\gg 1$. 

To keep things simple in this demonstration, we also neglect cooperativity. Now instead of (\ref{zzz}) we find 
\BEA
&&{\cal Z}_{\Nhb}=\frac{1}{n!}V^n Z_{\Nhb},\nonumber\\
&&Z_{\Nhb}=1+\sum_{\ell=1}^{4\Nhb} \ell ! \binom{n}{\ell}
\binom{4\Nhb}{\ell}
V^{-\ell}
v^\ell e^{-\beta\ell \varepsilon}.
\label{kayur}
\EEA
Assuming $n\gg 1$ (but still $\Nhb>n$) does not allow us to approximate uniformly in (\ref{kayur}) 
$\ell ! \binom{n}{\ell}\simeq n^\ell$, and hide $n$ and $V$ in the concentration parameter $n/V$, which is the essence of the grand canonical ensemble. Using the confluent hypergeometric function ${\cal U}(x,y,z)$ we get
\BEA
\label{german1}
Z_{\Nhb}=(-1)^n \Big(\frac{ve^{-\beta\varepsilon}}{V}\Big)^n
{\cal U}\Big(-n,1-n+4\Nhb,-\frac{Ve^{\beta\varepsilon}}{v}\Big).
\EEA
Employing a known formula
\BEA
\frac{\partial {\cal U}(x,y,z) }{\partial z} =-x\,{\cal U}(x+1,y+1,z),
\EEA
we find
\BEA
\frac{\overline{n_b}}{n}
=\frac{1}{n}v\frac{\partial}{\partial v}\ln Z_{\Nhb}
=1+\frac{{\cal U}(1-n,2-n+4\Nhb,-\frac{Ve^{\beta\varepsilon}}{v})
\frac{Ve^{\beta\varepsilon}}{v}
}{{\cal U}(-n,1-n+4\Nhb,-\frac{Ve^{\beta\varepsilon}}{v})}.
\label{dedal}
\EEA
Once $\overline{n_b}$ is the main quantity of the grand canonical ensemble, (\ref{dedal}) can be used to compare the present calculation with this ensemble.
In the limit $\Nhb>n\gg 1$, we now show that neither $\overline{n_b}$ nor $\frac{\overline{n_b}}{n}$ depends on $n$ and $V$ only via $n/V$ (this feature is the hallmark of the grand canonical ensemble). 
We have from (\ref{dedal}):
\BEA
\label{otark}
&&\frac{\overline{n_b}}{n}(n=10^4, \Nhb=2\times 10^4, V=10^4)=0.877,\\
&&\frac{\overline{n_b}}{n}(n=2\times 10^4, \Nhb=2\times 10^4, V=2\times 10^4)=0.764,
\label{otark2}
\EEA
where in (\ref{otark2}) we increased $n$ and $V$ two times. So if $\frac{\overline{n_b}}{n}$ were a function of $n/V$ only, it had to stay constant. Eqs.~(\ref{otark}, \ref{otark2}) are also inconsistent with ${\overline{n_b}}$ being a function of $n/V$ only.

\end{document}